# A Practical Framework for ROI Detection in Medical Images - a case study for hip detection in anteroposterior pelvic radiographs

Feng-Yu Liu[1], Chih-Chi Chen[2], Shann-Ching Chen[1,*], Chien-Hung Liao[3,4,*]

[1]Compal Electronics, Smart Device Business Group, Taipei, Taiwan
[2]Department of Physical Medicine and Rehabilitation, Chang Gung Memorial Hospital, Linkou, Chang Gung University, Taoyuan, Taiwan
[3]Department of Trauma and Emergency Surgery, Chang Gung Memorial Hospital, Linkou, Chang Gung University, Taoyuan, Taiwan
[4]Center for Artificial Intelligence in Medicine, Chang Gung Memorial Hospital, Linkou, Taoyuan, Taiwan

[*]Corresponding Authors: Shann-Ching Chen: ShannC_Chen@compal.com, Chien-Hung Liao: surgymet@gmail.com

## 1. Abstract

### 1.1 Purpose
Automated detection of region of interest (ROI) is a critical step for many medical image applications such as heart ROIs detection in perfusion MRI images, lung boundary detection in chest X-rays, and femoral head detection in pelvic radiographs. Thus, we proposed a practical framework of ROIs detection in medical images, with a case study for hip detection in anteroposterior (AP) pelvic radiographs.

### 1.2 Materials and Methods
We conducted a retrospective study which analyzed hip joints seen on 7,399 AP pelvic radiographs from three diverse sources, including 4,290 high resolution radiographs from Chang Gung Memorial Hospital Osteoarthritis, 3,008 low to medium resolution radiographs from Osteoarthritis Initiative, and 101 heterogeneous radiographs from Google image search engine. We presented a deep learning-based ROI detection framework utilizing single-shot multi-box detector (SSD) with ResNet-101 backbone and customized head structure based on the characteristics of the obtained datasets, whose ground truths were labeled by non-medical annotators in a simple graphical interface.

### 1.3 Results
Our method achieved average intersection over union (IoU)=0.8115, average confidence=0.9812, and average precision with threshold IoU=0.5 ($AP_{50}$)=0.9901 in the independent test set, suggesting that the detected hip regions have appropriately covered main features of the hip joints.

### 1.4 Conclusion
The proposed approach featured on low-cost labeling, data-driven model design, and heterogeneous data testing. We have demonstrated the feasibility of training a robust hip region detector for AP pelvic radiographs. This practical framework has a promising potential for a wide range of medical image applications.

## 2. Introduction

The deep convolutional neural network (DCNN) has shown a significant breakthrough in many aspects of commercial image differentiation and identification. In recent years, DCNNs also played important roles in medical image analysis. For example, the ChestX-ray8 ([1](#)) and MURA ([2](#)) are two representative studies utilizing the state-of-the-art DCNN classification and visualization models to detect and locate disease patterns in chest and musculoskeletal radiographs. The advantage of this straightforward "one-step" classification strategy is to identify abnormal and large features that require follow up attention. This approach has been applied to many medical image applications, including assistant diagnosis of hip fracture ([3](#)) and osteoarthritis ([4](#)) in AP pelvic radiographs.

Furthermore, some studies employ a more delicate "two-step" classification strategy, which first detects specific ROIs ([5-9](#)) followed by conventional classification methods ([10-12](#)). A seminal work is the automatic knee osteoarthritis diagnosis in lateral knee radiographs, where knee regions are first identified ([13](#)) followed by classification and heatmap visualization ([14](#)). The advantage of this "two-step" approach is the capability to identify more subtle localized abnormalities and has gradually became the mainstream technology, especially for the analysis of AP pelvic radiographs including fracture subclass identification ([15](#)), hip osteoarthritis grading ([16](#)), and avascular necrosis detection ([17](#)).

A critical component for successful "two-step" classification system is accurate ROI detection, which falls into computer vision object detection task ([18](#)) usually tackled by different strategies including bounding-box regression, pixel-level segmentation and landmark detection based approach. Among these methods, the bounding-box based methodology is advantageous for its cheap annotation cost and simple implementation, which is proven to be effective in popular computer vision applications such as human detection in surveillance videos and real-time pedestrian and vehicle detection in video. In order to identify multiple objects across different scales in one image, it is important to generate anchor boxes of varied sizes and aspect ratios for hyper-parameter optimization. However, there are usually a small number of non-overlapping objects in medical images, for example only one heart ROI in lung X-rays and two hip ROIs in hip radiographs. It is not optimal to apply the same object detection parameters on underlying different applications.

In this work, we propose a practical framework of ROI detection and parameter selection in medical images. To the best of our knowledge, this is the first work that provides a systematic guideline for parameter selection and has a promising potential for a wide range of medical image applications.

## 3. Materials and Methods

### 3.1 Dataset Acquisition

This retrospective study analyzed hip joints seen on 7,399 anteroposterior pelvic radiographs from three diverse sources, including the first Chang Gung Memorial Hospital Osteoarthritis (CGOA) dataset containing 4,290 high resolution radiographs, the second Osteoarthritis Initiative Hip (OAIH, pelvic radiograph dataset extracted from subset of data from the OAI ([19](#))) dataset containing 3,008 radiographs with relatively lower resolutions, and the third Google Image Search (GIS) dataset containing 101 heterogeneous radiographs. Table 1 lists the summary statistics of these datasets. This experimental design which utilizes radiographs generated from diverse sources of different imaging protocols, resolutions and ethnicities ensures that model generalization can be achieved. Details of these three datasets can be found in Appendix E1 (supplement).

| Datasets | Number of images | Max (pixels) | Min (pixels) | Median (pixels) | Mean (pixels) | Standard Deviation | Recruit Year | Hip Region Annotation | Etiology Reading |
|---|---|---|---|---|---|---|---|---|---|
| CGOA | 4290 | 4280 | 1616 | 2688 | 2635.8 | 201.1 | 2008-2017 | RL, SC | CL |
| OAIH | 3008 | 1080 | 466 | 535 | 571.3 | 97.0 | 2004-2014 | SI | N/A |
| GIS | 101 | 4256 | 225 | 258 | 515.3 | 626.6 | N/A | SC | N/A |

**Table 1:** Summary statistics of the three datasets used in this study.

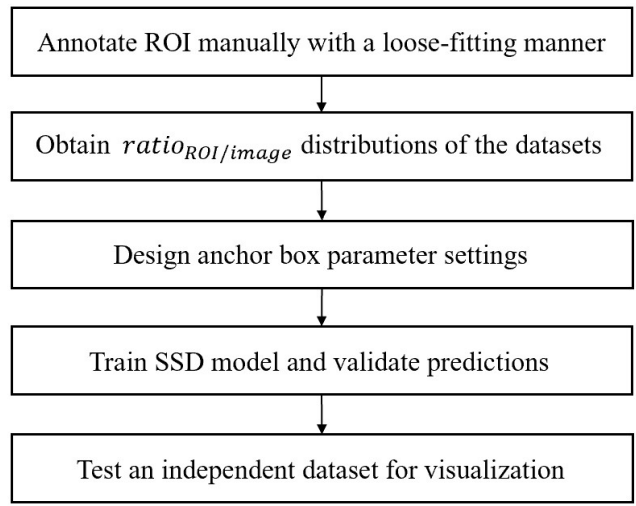

**Figure 1**: Overview of the proposed framework for hip ROI detection.

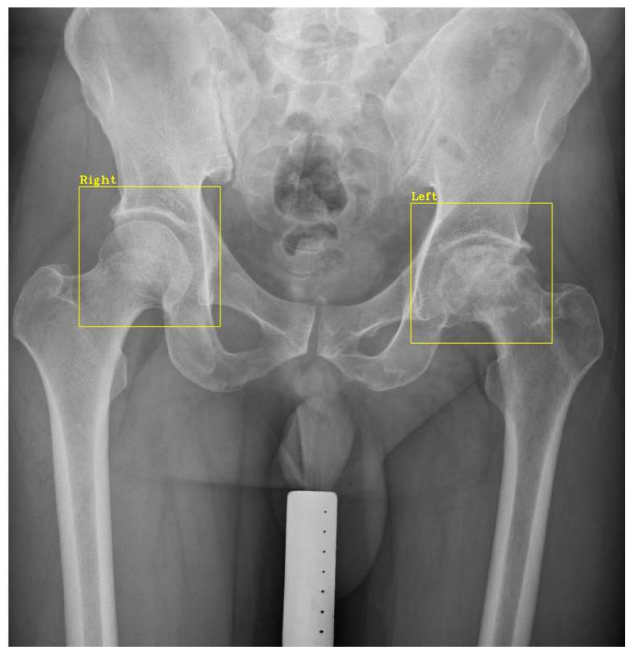

**Figure 2:** Manually-annotated hip regions of interest in an anteroposterior pelvic radiograph, where yellow bounding boxes indicate left and right hip locations.

### 3.2 Data Annotation
Figure 1 shows the overview of the proposed framework. To annotate hip regions of interests, we employed three annotators RL, SI and SC, who are data scientists with limited medical knowledge, to place square bounding boxes approximately centered at the femoral head or the artificial hip joint with a customized GUI software. Clinical readings for fracture identification were performed by one chief surgeon CL with 15 years of clinical experience in reading radiographs. The fracture labels are utilized for data exclusion criteria. Figure 2 shows a sample radiograph with yellow bounding boxes highlighting the locations of right and left hips. It is noted that identifying a complete round femoral head in healthy hips is relatively straightforward; however, for cases with severe osteoarthritis or other hip conditions with collapsed or abnormal femoral heads, we employ a loose-fitting manner to make sure every hip joint lays appropriately in the bounding box. As manual annotation by doctors is time consuming and usually the bottleneck for medical image analysis, the approximate identification of hip regions can be done by non-physicians in an effective and efficient manner.

### 3.3 Proposed SSD Model Architecture for ROI Detection in Hip Radiographs
The proposed hip region detection architecture simplifies existing SSD model architecture (Figure 3) ([9]) which was originally developed for detecting multiple objects with different sizes and aspect ratios in applications such as video surveillance, robot vision, and traffic monitoring. In these applications, multiple objects with various sizes and shapes within an image are needed to be detected, where the SSD networks utilized multiple features layers as convolutional predictors with different scales and aspect ratios for the anchor boxes (Figure 3A). The parameter selection is usually empirical and redundant depending on the available datasets and applications. For ROI detection in medical images, there is usually one or two important organs (e.g., one heart, two hips or two knees) in one radiograph. It is feasible to have a simplified SSD architecture with only one feature layer as the only convolutional predictor, with an appropriate receptive field size, one aspect ratio (1:1 in for hip ROI) and a small set of scales (e.g., 3 or 6 in Figure 3B). We replaced SSD VGG-16 backbone by ResNet-101 (11) backbone which was pre-trained on ImageNet (20). All these modifications could reduce ROI detections from several thousands to a few hundreds, decrease training time and complexity as well increase detection accuracy and confidence.

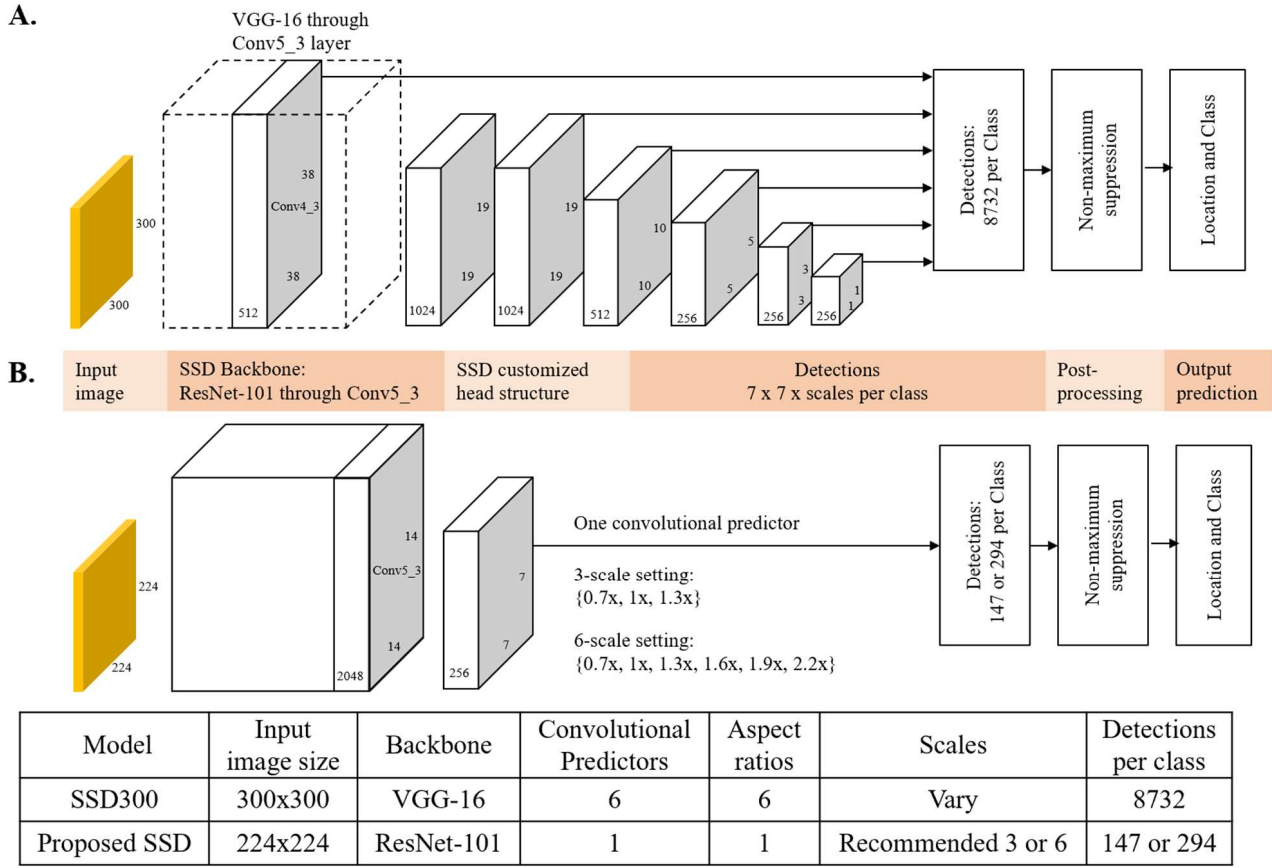

| Model | Input image size | Backbone | Convolutional Predictors | Aspect ratios | Scales | Detections per class |
|---|---|---|---|---|---|---|
| SSD300 | 300x300 | VGG-16 | 6 | 6 | Vary | 8732 |
| Proposed SSD | 224x224 | ResNet-101 | 1 | 1 | Recommended 3 or 6 | 147 or 294 |

**Figure 3:** Comparison of SSD model architectures. **(A)** Original SSD300 architecture with VGG-16 backbone. **(B)** Proposed architecture with ResNet-101 backbone and other customized settings.

To best determine the anchor box parameter settings, we first define $ratio_{ROI/image}$ as the size of the square ROI divided by the length of the long side of the input image (zero padding to a square if needed). This ratio is designed as a normalizer which makes the anchor boxes and ROI instances compatible across different datasets. Next, we analyzed image size distributions (Figure 4A) and $ratio_{ROI/image}$ distributions (Figure 4B) of the three available heterogeneous datasets, where the ratios lie mostly between 10% to 30%. We specified the input image size of 224x224 pixels split by 7x7 grid cells, where each grid cell is of size 32x32 pixels. We set 6 equal spaced scales parameters {0.7, 1.0, 1.3, 1.6, 1.9, 2.2} so that the smallest and largest anchor boxes could cover 10% and 31.4% of the images respectively. This design ensures that the designed anchor boxes can identify appropriate hip ROIs in the datasets.

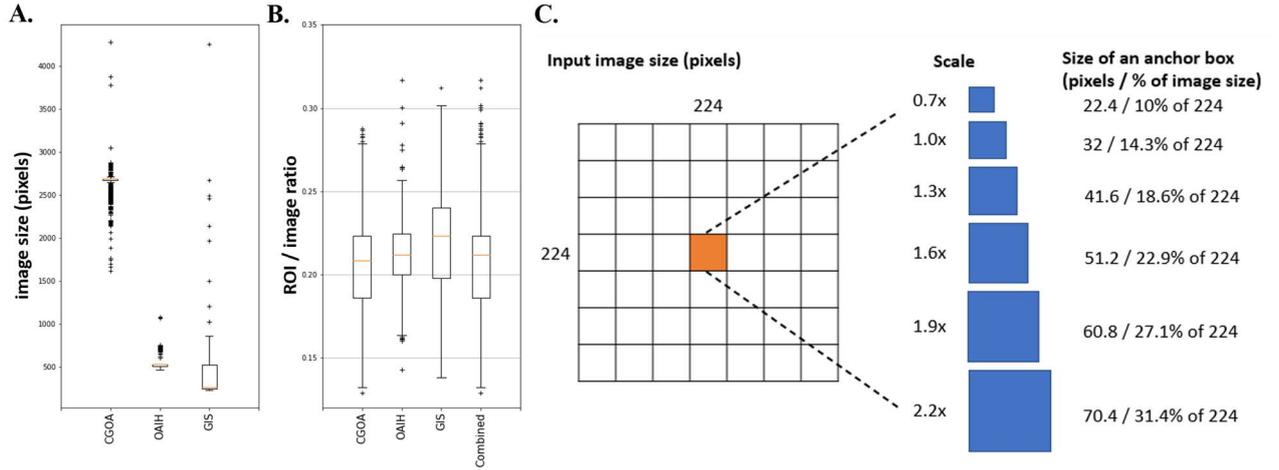

**Figure 4:** Comparison of three radiographic datasets distributions and generation of anchor boxes. **(A)** The image size distributions of the three datasets. **(B)** The $ratio_{ROI/image}$ distributions of three datasets. **(C)** Generation of anchor boxes for the one feature layer of the customized SSD head structure. With an input square image with 224x224 pixels, there are 7x7 grid cells with 32x32 pixels with scale=1, and each grid cell can use different scale parameters to generate various sizes of anchor boxes covering 10% to 31.4% of the input image size, depending on the training image size distributions.

### 3.4 Data Preprocessing, Training, and Evaluation

For data preprocessing, each radiograph was zero padding to a square image and resized to 224x224 pixels with 8-bit grayscale before feeding into the model. The model was implemented by fastai v0.7 library (21) with Python 3.6.4, and we randomly split the combined CGOA and OAIH dataset into 90% for training and 10% for validation, and used all 101 GIS radiographs as the independent test dataset (more training details in Appendix E2). For evaluation, we used the standard IoU metric for comparing the predicted bounding box $B_{pred}$ and ground truth bounding box $B_{gt}$:

$$IoU = \frac{B_{pred} \cap B_{gt}}{B_{pred} \cup B_{gt}}$$

where ∩ and ∪ denote intersection and union respectively. We reported the associated confidence for each predicted bounding box, average IoU, average confidence, minimal confidence and $AP_{50}$, as the 0.5 cutoff indicate poor ROI detection which may cause issues for downstream analysis.

### 4. Results
### 4.1 Demographics of the study population

Among the first CGOA dataset containing 4,290 high resolution radiographs, the average age of 2660 trauma patients (40.8% male and ranged in age from 18-102) included in the study was 63.06 ± 15.72 (standard deviation), and the average age of 1630 'normal' cases was 44.88 ± 20.46 (68.2% male and ranged in age from 18-88) from emergency room without undergoing hip surgery. The age and gender distribution of this dataset are shown in Appendix E3. The second OAIH dataset is a consolidated pelvic radiograph dataset extracted from subset of data from the OAI project, which recruited 4,796 participants from February 2004 to May 2006 to form a baseline cohort (58% female and ranged in age from 45-79 at time of recruitment). The third GIS dataset was acquired through Google image search engine (keywords included "pelvic radiograph", "pelvis xray" and "hip osteoarthritis xray") and the demographics is not available.

**A.**

| Experiment | Training and validation sets | Testing set | Anchor box parameter settings |
|---|---|---|---|
| Exp1 | CGOA+OAIH | GIS | Layer[4,2,1] and 3-scale setting with range(start=0.7, stop=1.3, step=0.3) |
| Exp2 | CGOA+OAIH | GIS | Layer[7] and 3-scale setting with range(start=0.7, stop=2.2, step=0.75) |
| Exp3 | CGOA+OAIH | GIS | Layer[7] and 6-scale setting with range(start=0.7, stop=2.2, step=0.3) |
| Exp4 | CGOA+OAIH | GIS | Layer[7] and 16-scale setting with range(start=0.7, stop=2.2, step=0.1) |
| Exp5 | CGOA+OAIH | GIS | Layer[7] and 31-scale setting with range(start=0.7, stop=2.2, step=0.05) |
| Exp6 | CGOA+OAIH | GIS | Layer[14] and 31-scale setting with range(start=0.7, stop=2.2, step=0.05) |
| Exp7 | CGOA | OAIH | Layer[7] and 6-scale setting with range(start=0.7, stop=2.2, step=0.3) |
| Exp8 | CGOA | GIS | Layer[7] and 6-scale setting with range(start=0.7, stop=2.2, step=0.3) |
| Exp9 | OAIH | CGOA | Layer[7] and 6-scale setting with range(start=0.7, stop=2.2, step=0.3) |
| Exp10 | OAIH | GIS | Layer[7] and 6-scale setting with range(start=0.7, stop=2.2, step=0.3) |

**B.**

| Experiment | Validation | | Test | | | |
|---|---|---|---|---|---|---|
| | avg IoU | avg confidence | avg IoU | avg confidence | minimal IoU | $AP_{50}$ |
| Exp1 | 0.8477 | 0.6932 | 0.7912 | 0.7130 | 0.4042 | 0.9851 |
| Exp2 | 0.8565 | 0.9075 | 0.8105 | 0.9422 | 0.3829 | 0.9901 |
| **Exp3** | **0.8571** | **0.9582** | **0.8115** | **0.9812** | **0.3861** | **0.9901** |
| Exp4 | 0.8471 | 0.9644 | 0.8011 | 0.9843 | 0.3809 | 0.9950 |
| Exp5 | 0.8432 | 0.9403 | 0.7749 | 0.9751 | 0.3680 | 0.9901 |
| Exp6 | 0.8267 | 0.9892 | 0.7918 | 0.9950 | 0.3732 | 0.9950 |
| Exp7 | 0.8460 | 0.9496 | 0.8307 | 0.9428 | 0.3622 | 0.9945 |
| Exp8 | 0.8460 | 0.9496 | 0.7951 | 0.9634 | 0.3652 | 0.9851 |
| Exp9 | 0.8664 | 0.8971 | 0.7633 | 0.8667 | 0.3075 | 0.9846 |
| Exp10 | 0.8664 | 0.8971 | 0.7813 | 0.9221 | 0.3449 | 0.9901 |

**Table 2:** Evaluation of model performance. **(A)** Examination of different training datasets and anchor box parameter settings. **(B)** Model performance of validation and test sets.

### 4.2 Model Settings and Performance

According to the dataset distributions examined in Figure 4, we evaluated various training and validation datasets with different anchor box parameter settings in Table 2A. With the input image $I$ with 224x224 pixels, Layer[7] used in Exp2-7 indicates that one convolutional predictor splits $I$ into 7x7 grid cells of 32x32 pixels (224/7=32), while Layer[4,2,1] used in Exp1 indicates that three feature layers split $I$ into (a) 4x4 grid cells of 56x56 pixels (224/4=56), (b) 2x2 grid cells of 112 x 112 pixels (224/2=112), and (c) 1x1 grid cell of 224x224 pixels (224/1=224), respectively. The scale combination applied on the anchor grid is denoted as, for example, range(start=0.7, stop=1.3, step=0.3), which stands for the 3-scale setting {0.7, 1.0, 1.3}.

We first evaluated the default fastai SSD parameters for the PASCAL Visual Object Classes dataset ([22]) where parameters are optimal for recognizing objects from a number of visual object classes in realistic scenes. As shown in Table 2B, Exp1 achieved average IoU=0.8477 and average confidence=0.6932 on the validation set and average IoU=0.7912 and average confidence=0.7130 on independent GIS testing set. In order to achieve better performance, we adjusted the parameters as Layer[7] and Layer[14] with various scale settings (Exp2-6). The best performance was achieved in Exp3 with average IoU=0.8571 and average confidence=0.9582 on the validation set and average IoU=0.8115, average confidence=0.9812, minimal IoU=0.3861 and $AP_{50}$=0.9901 on the testing set. These results are quite promising with extremely high average confidence both on the validation set and the totally independent GIS test set. We further examined the performance of optimal Exp3 parameters only with COGA or OAIH in the training and validation sets (Exp7-10). The last 4 rows in Table 2B indicates that without a more diverse training set covering high-resolution radiographs (CGOA) and low to medium resolution radiographs (OAIH), the average IoU and confidence are generally lower in both validation and test sets. Therefore, it is desirable to have heterogeneous training, validation and test datasets for model generalization and performance evaluation.

| Datasets | Number of images | Number of hip ROIs | Avg IoU | Avg Confidence | Minimal IoU | Number of hip ROIs with IoU<0.5 | AP$_{50}$ |
|---|---|---|---|---|---|---|---|
| All: CGOA & OAIH & GIS | 7399 | 14798 | 0.9176 | 0.9688 | 0.3861 | 2 | 0.9999 |
| Train: 90% CGOA & OAIH | 6568 | 13136 | 0.9260 | 0.9698 | 0.5955 | 0 | 1 |
| Valid: 10% CGOA & OAIH | 730 | 1460 | 0.8571 | 0.9582 | 0.5907 | 0 | 1 |
| Test: GIS | 101 | 202 | 0.8115 | 0.9812 | 0.3861 | 2 | 0.9901 |

**Table 3:** Detailed performance metrics with the optimal parameters using the proposed hip region detection architecture.

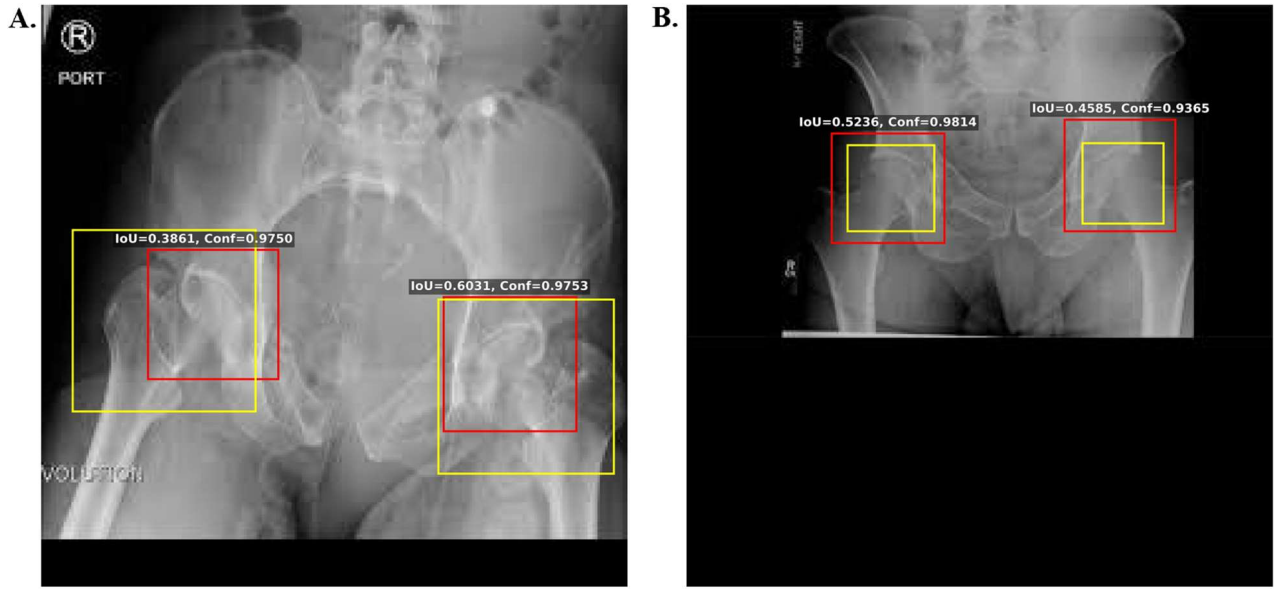

**Figure 5:** Two radiographs from the test dataset with IoU < 0.5. Yellow boxes indicate manual labels and red boxes indicate detected hip ROIs. **(A)** The radiograph with the lowest IoU (0.3861) on the right hip ROI, where the manual annotation is slightly larger than the inferred ROI. **(B)** The radiograph with the second lowest IoU (0.4585) on the left hip ROI, where the manual annotation is slightly smaller than the inferred ROI.

### 4.3 Visualization

In Table 3, we take a closer look at the best performance results and carefully examined those cases where hip ROIs with IoU<0.5. As AP$_{50}$ metrics are 1 in both training and validation set and 0.9901 in the independent GIS test set, we only identified two cases below IoU 0.5 cutoff which may indicate poor ROI detection and cause issues for downstream analysis. As shown in Figure 5A, the radiograph has the lowest IoU (0.3861) on the right hip ROI, where the manual annotation is slightly larger than others and the inferred ROI has moderate size and covers most key features of the hip area. In Figure 5B, the radiograph has the second lowest IoU (0.4585) on the left hip ROI, where the manual annotation is slightly smaller than others and the inferred ROI has moderate size and covers most key features of the hip area. When including these two cases, AP$_{50}$ on the test set is 0.9901 and the visual examination suggested that the inferred results are more consistent across three different data sources and may be more suitable for automated downstream analysis.

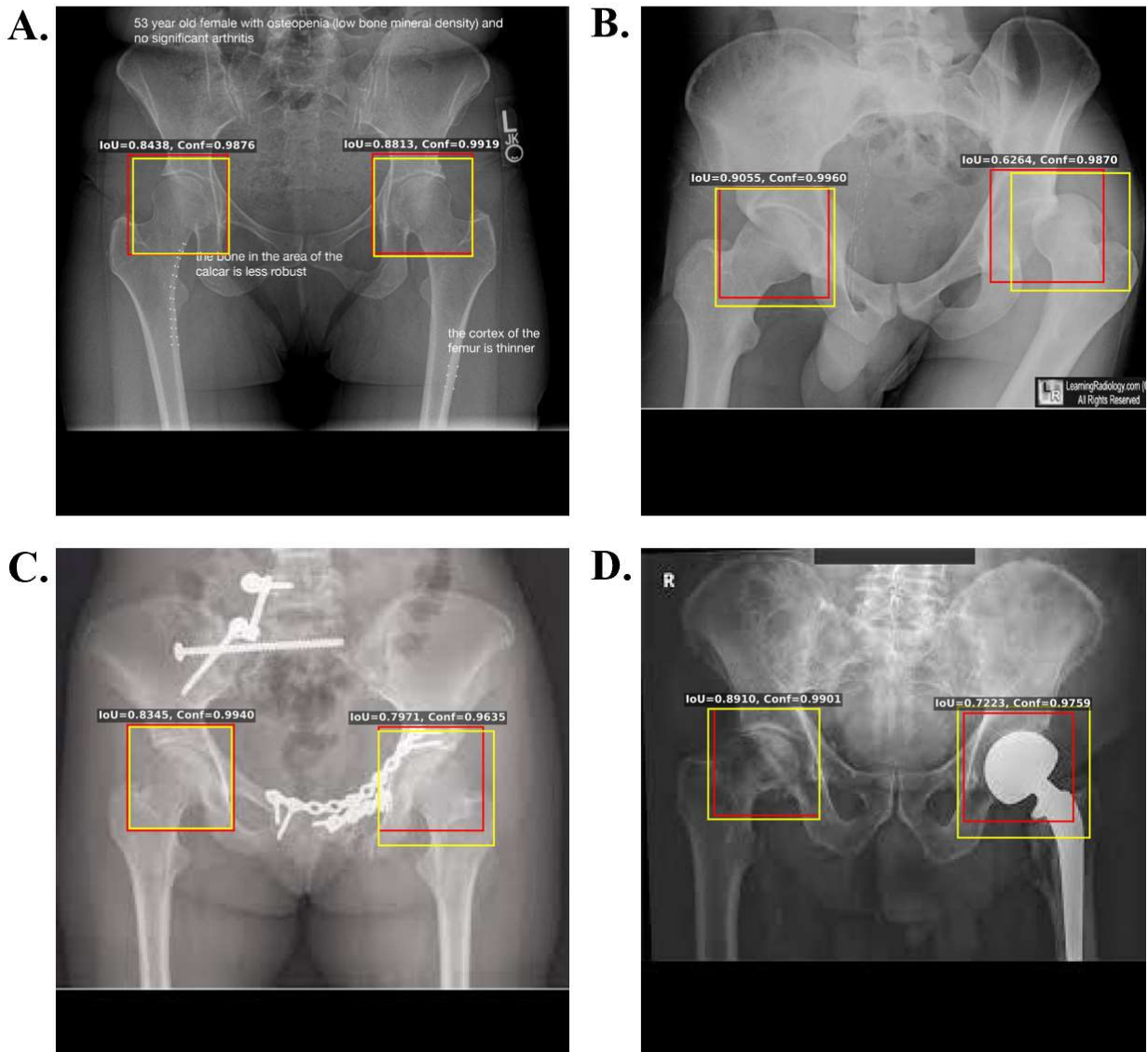

**Figure 6.** Example hip ROI detection results on the Google Image Search dataset. Yellow boxes indicate manual labels and red boxes indicate detected hip ROIs. **(A)** A radiograph with some text outside the key hip area. **(B)** A radiograph with dislocation on the left hip but the detected hip ROI covers most key features of the hip area. **(C)** A radiograph with chains on the left hip but still can be detected correctly. **(D)** A radiograph with artificial joint on the left hip but still can be detected correctly.

We further examined other radiographs in the heterogeneous test set and Figure 6 showed several representative results. Figure 6A shows a radiograph with some text outside the key hip area. Figure 6B shows a radiograph with dislocation on the left hip but the detected hip ROI covers most key features of the hip area. Figure 6C shows a radiograph with chains on the left hip but still can be detected correctly. Figure 6D shows a radiograph with artificial joint on the left hip but still can be detected correctly. These results suggested that our model with specially designed anchors and trained by diverse datasets is a general and robust hip region detector which can be applicable for a wide range of heterogeneous datasets with different qualities and resolutions and can be potentially useful for automated assessment of many hip bone conditions. We provided visualization of all 101 GIS hip ROI detection results in Appendix E4.

## 5. Discussion

In this work, we have demonstrated a practical framework for detecting regions of interest in medical images. With the case study for hip detection in anteroposterior pelvic radiographs, we achieved average IoU over 80% and average confidence higher than 95%. These independent test set shows promising ROI detection results on Google Image Search images with heterogenous resolutions and appearance. The manual annotation with approximation identification of hip regions that can be performed by non-physicians in an effective and inexpensive manner.

Comparing to traditional object detection tasks which need to recognize multiple objects with different sizes and aspect ratios in images and videos, the proposed SSD architecture has the advantages of simpler structure, higher IoU accuracy and reliable confidence. The challenge of determining those empirical parameter settings now relies on the basic statistics on the available datasets to generate enough anchor boxes. Our results suggest that more anchors do not necessarily encourage higher IoU but may decrease the prediction performance. The proposed method provides a more effective approach for anchor design and parameter optimization.

Automated and accurate ROI detection is critical for automated computer assisted analysis for screening and diagnostics. The proposed framework provides a guideline for parameter settings in anchor-based object detection algorithms, and it is especially useful for applications such as joint identification in orthopedics radiographs and other medical image problems. However, there are still some limitations in the existing method. First, the manual hip annotation with loose-fitting criteria is not unique and can be varied from person to person, especially for those cases with severe osteoarthritis or other hip conditions with collapsed or abnormal femoral heads. As a result, a detected ROI with higher IoU usually indicates goodness of such detection, while a detected ROI with lower IoU could be possibly caused by inconsistent human annotations rather than an inappropriate ROI is detected, and a closer visual examination is needed. Second, the utility of such ROI detection approach highly depends on the downstream applications. For hip detection in anteroposterior pelvic radiographs, this automated hip ROI detection enables applications such as surgical prediction of total hip replacement, fracture identification, osteoarthritis assessment, osteoporosis and avascular necrosis or other hip conditions. The evaluations of such applications and integrated system remain to be investigated in future works. Such systems need to integrate clinical information, which differs from clinical practice considerations and remain to be open research topics. Lastly, more validation datasets from hospitals should be further investigated in the real world applications.

Medical AI is progressive to change the healthcare system, and various DCNNs showed that it is feasible to detect the lesion from pathologic images, electrocardiography, and radiography. These algorithms presented outstanding achievement in disease detection or prediction whose performance is not inferior to the physicians. These results inspire us that DCNN might help the human in the healthcare sector in different ways. However, the development of medical AI is not accessible due to some limitations. The data clearance and accurate label were considered fundamental for deep learning because of the limited size of medical images and the high cost of a medical expert to perform labeling. Moreover, the hip ROI detection system can help the physician to label the lesion with a weak supervision way, where we can pick out the hip regions and save time for the physician to crop and copy the images. To reduce the barrier between an outliner and attract more physicians and scientists to join a new raising technologic field is another issue in the real world. In this study, we developed the diagnostic assistance system and created a useful tool for reducing the workload during data collection and tuning.

In conclusion, with the proposed DCNN framework, we can identify the hip joint with high accuracy, reliability, and reproducibility.

## 6. Abbreviations

ROI = region of interest, IoU = intersection over union,
SSD = single-shot multi-box detector, DCNN = deep convolutional neural network,
OAI = Osteoarthritis Initiative, OAIH = Osteoarthritis Initiative Hip dataset,
CGMH = Chang Gung Memorial Hospital, CGOA = Chang Gung Memorial Hospital Osteoarthritis dataset, GIS = Google image search dataset.

## 7. Summary

A specialized DCNN model was trained to detect hip region of interests from radiographs, with average IoU=0.8115, average confidence=0.9812, and average precision with threshold IoU=0.5 0.9901 in the independent Google image search test set. These results and visualization suggested that detected hip ROIs are accurate and useful for suitable for automated downstream analysis.

## 8. Key points

- In this study, we proposed a practical approach for ROI detection in medical image applications, with a case study on for hip detection in anteroposterior pelvic radiographs.
- The proposed approach featured on low-cost labeling, data-driven model design, and heterogeneous data testing. We have demonstrated the feasibility of training a robust hip region detector for anteroposterior pelvic radiographs.
- This practical framework provided a systematic guideline for parameter selection and has a promising potential for a wide range of medical image applications.

**Additional Information**

Author contributions: F.-Y.L., S.-C.C. and C.-H.L. designed the experiments; C.-C.C. and C.-H.L. acquired radiographics for use in the study and provided strategic support; F.-Y.L. and S.-C.C. wrote code to achieve different tasks and carried out all experiments; F.-Y.L. implemented the annotation tools for data annotation; F.-Y.L., S.-C.C. and C.-H.L. provided labels for use in measuring algorithm performance; F.-Y.L. drafted the manuscript; S.-C.C. helped extensively with writing the manuscript; S.-C.C. and C.-H.L supervised the project; All authors read and approved the final manuscript.

Competing interests: F.-Y.L. and S.-C.C. are employees of Compal Electronics. C.-C.C. and C.-H.L. declare no relationships with any companies, whose products or services may be related to the subject matter of the article.

Code availability: Available upon request from the authors.

Data availability: The CGOA dataset is not publicly available due to restrictions in the data sharing agreements with the Chang Gung Memorial Hospital Institutional Review Board (IRB). The OAI and OAIH datasets are publicly available by the Osteoarthritis Initiative website resources: https://nda.nih.gov/oai. The GIS dataset is available by using the Google Image Search engine.

Supplemental material is available for this article. **(Appendix E1-E4)**

## 9. References

1. Wang X, Peng Y, Lu L, Lu Z, Bagheri M, Summers RM, editors. Chestx-ray8: Hospital-scale chest x-ray database and benchmarks on weakly-supervised classification and localization of common thorax diseases. Proceedings of the IEEE conference on computer vision and pattern recognition; 2017.

2. Rajpurkar P, Irvin J, Bagul A, Ding D, Duan T, Mehta H, et al. Mura: Large dataset for abnormality detection in musculoskeletal radiographs. arXiv preprint arXiv:171206957. 2017.

3. Cheng C-T, Ho T-Y, Lee T-Y, Chang C-C, Chou C-C, Chen C-C, et al. Application of a deep

learning algorithm for detection and visualization of hip fractures on plain pelvic radiographs. European radiology. 2019;29(10):5469-77.

4. Xue Y, Zhang R, Deng Y, Chen K, Jiang T. A preliminary examination of the diagnostic value of deep learning in hip osteoarthritis. PLoS One. 2017;12(6):e0178992.

5. Redmon J, Divvala S, Girshick R, Farhadi A, editors. You only look once: Unified, real-time object detection. Proceedings of the IEEE conference on computer vision and pattern recognition; 2016.

6. Redmon J, Farhadi A, editors. YOLO9000: better, faster, stronger. Proceedings of the IEEE conference on computer vision and pattern recognition; 2017.

7. Redmon J, Farhadi A. Yolov3: An incremental improvement. arXiv preprint arXiv:180402767. 2018.

8. Bochkovskiy A, Wang C-Y, Liao H-YM. YOLOv4: Optimal Speed and Accuracy of Object Detection. arXiv preprint arXiv:200410934. 2020.

9. Liu W, Anguelov D, Erhan D, Szegedy C, Reed S, Fu C-Y, et al., editors. Ssd: Single shot multibox detector. European conference on computer vision; 2016: Springer.

10. Simonyan K, Zisserman A. Very deep convolutional networks for large-scale image recognition. arXiv preprint arXiv:14091556. 2014.

11. He K, Zhang X, Ren S, Sun J, editors. Deep residual learning for image recognition. Proceedings of the IEEE conference on computer vision and pattern recognition; 2016.

12. Xie S, Girshick R, Dollár P, Tu Z, He K, editors. Aggregated residual transformations for deep neural networks. Proceedings of the IEEE conference on computer vision and pattern recognition; 2017.

13. Tiulpin A, Thevenot J, Rahtu E, Saarakkala S, editors. A novel method for automatic localization of joint area on knee plain radiographs. Scandinavian Conference on Image Analysis; 2017: Springer.

14. Tiulpin A, Thevenot J, Rahtu E, Lehenkari P, Saarakkala S. Automatic knee osteoarthritis diagnosis from plain radiographs: a deep learning-based approach. Scientific reports. 2018;8(1):1-10.

15. Krogue JD, Cheng KV, Hwang KM, Toogood P, Meinberg EG, Geiger EJ, et al. Automatic Hip Fracture Identification and Functional Subclassification with Deep Learning. Radiology: Artificial Intelligence. 2020;2(2):e190023.

16. von Schacky CE, Sohn JH, Liu F, Ozhinsky E, Jungmann PM, Nardo L, et al. Development and Validation of a Multitask Deep Learning Model for Severity Grading of Hip Osteoarthritis Features on Radiographs. Radiology. 2020;295(1):136-45.

17. Li Y, Li Y, Tian H. Deep Learning-based End-to-end Diagnosis System for Avascular Necrosis of Femoral Head. arXiv preprint arXiv:200205536. 2020.

18. Zhao Z-Q, Zheng P, Xu S-t, Wu X. Object detection with deep learning: A review. IEEE transactions on neural networks and learning systems. 2019;30(11):3212-32.

19. Joseph G, Hilton J, Jungmann P, Lynch J, Lane NE, Liu F, et al. Do persons with asymmetric hip pain or radiographic hip OA have worse pain and structure outcomes in the knee opposite the more affected hip? Data from the Osteoarthritis Initiative. Osteoarthritis and cartilage. 2016;24(3):427-35.

20. Russakovsky O, Deng J, Su H, Krause J, Satheesh S, Ma S, et al. Imagenet large scale visual recognition challenge. International journal of computer vision. 2015;115(3):211-52.

21. Howard J, Gugger S. Fastai: A layered API for deep learning. Information. 2020;11(2):108.

22.	Everingham M, Williams CK. The PASCAL visual object classes challenge 2007 (VOC2007) results.

23.	Lin T-Y, Goyal P, Girshick R, He K, Dollár P, editors. Focal loss for dense object detection. Proceedings of the IEEE international conference on computer vision; 2017.

24.	Kingma DP, Ba J. Adam: A method for stochastic optimization. arXiv preprint arXiv:14126980. 2014.

25.	Smith LN, editor Cyclical learning rates for training neural networks. 2017 IEEE Winter Conference on Applications of Computer Vision (WACV); 2017: IEEE.

26.	Smith LN, Topin N, editors. Super-convergence: Very fast training of neural networks using large learning rates. Artificial Intelligence and Machine Learning for Multi-Domain Operations Applications; 2019: International Society for Optics and Photonics.

**Appendix E1**
**Dataset Acquisition**
The first CGOA dataset utilized the Chang Gung Trauma Registry Program in Chang Gung Memorial Hospital (CGMH), Linkou, Taiwan. As a Level I trauma center and the largest Taiwanese medical center, CGMH network has a total of 10,050 beds. Demographic data, medical data, perioperative procedures, hospital procedures, medical imaging findings, follow-up data and information regarding complications were recorded prospectively in CGMH health information system. We extracted the AP pelvic radiographs and medical records of all 2660 trauma patients (40.8% male and ranged in age from 18-102) treated at CGMH from August 2008 to December 2017. This CGOA data collection is consisted of 3,013 pairs of radiographs before and after total hip replacement surgery, of which 353 are fracture cases and excluded from the analysis, and 1,630 'normal' cases (68.2% male and ranged in age from 18-88) from emergency room without undergoing hip surgery. This gives us a total of 4,290 high-resolution radiographs with 8-bit color depth and sizes ranging from 1,616 x 732 pixels to 4,280 x 3,020 pixels. For deidentification purpose, each image file name is converted to a serial number and the upper and lower portions of the images containing patient privacy information such as names, medical record number and date of birth were cropped out. The Internal Review Board of CGMH approved this study.

The second OAIH dataset is a consolidated pelvic radiograph dataset extracted from subset of data from the OAI, a prospective, longitudinal, and observational study of knee osteoarthritis ([19](#)). The OAI project recruited 4,796 participants from February 2004 to May 2006 to form a baseline cohort (58% female and ranged in age from 45-79 at time of recruitment), and have retained most patients for follow up visits in clinic with either biospecimen collection and/or imaging at full-limb, hand, knee, or pelvis at 12-month, 24-month, 36-month, and 48-month, 72-month, and 96-month. Custom python scripts were implemented to extract images with suffix "_1x1.jpg" and semi-automatically categorized these images into hip versus non-hip images. In total we identified 11,354 AP pelvic radiographs (4,707 at baseline; 64 at 12-month visit; 3,638 at 48-month visit; 2,944 at 96-month visit), and we annotated hip locations from 3,008 radiographs (64 and 2,944 images from 12-month and 96-month visits respectively) with 8-bit color depth and sizes ranging from 466 x 431 pixels to 1,080 x 890 pixels (relatively lower resolutions).

The third GIS dataset was acquired through Google image search engine (keywords included "pelvic radiograph", "pelvis xray" and "hip osteoarthritis xray"). After manual curation to filter out irrelevant or poor-quality images such as images only containing one side of hip or no hips, non-xray images, images from animals such as dogs or cats, saturated images or out-of-focus images or the long side of the image less with 255 pixels, we collected 121 radiographs with a wide size distribution ranging from 225 x 225 pixels to 4,256 x 3,495 pixels. We further excluded 20 images abnormal images, (e.g., images from young kids less than or equal to three years old with different hip morphology, or images with red marker lines, green arrows, or occluding objects such as bandages or scissors near or on the hip regions), resulting 101 heterogenous radiographs as independent test data for the study.

**Appendix E2**
**Model Training**
The model was implemented by fastai v0.7 library ([21](#)) with Python 3.6.4, PyTorch v.0.4.1 and CUDA 9.0 on Ubuntu 16.04 operating system with one Nvidia 1080Ti GPU. We used focal loss with ($\alpha$=0.25, $\gamma$=5) ([23](#)) and Adam optimizer with ($\beta_1$=0.9, $\beta_2$=0.99) ([24](#)). We trained the model with minibatches of size 16 and utilized cyclical learning rate ([25](#)) and one cycle policy ([26](#)), where we first trained the last layer with learning rate lr=0.01, learning array lrs {lr/100, lr/10, lr} with 40 cycles, then trained the last two layers with a smaller learning rate lr=0.0025 with 40 cycles, and followed with unfreezing all layers with smaller learning rate lr=0.0025 with 40 cycles. We augmented the data during training by applying random rotations of up to 3 degrees with 0.9 probability and random lighting with (balance=0.5, contrast=0.5). We randomly split the combined CGOA and OAIH dataset into 90% for training and 10% for validation, and used all 101 GIS radiographs as the independent test dataset.

**Appendix E3**
**Characteristics of the Chang Gung Memorial Hospital Osteoarthritis (CGOA) dataset containing 4,290 high resolution radiographs**

|  | OA | Normal | p-value |
|---|---|---|---|
| Image numbers | 2660 | 1630 |  |
| Age | 63.06 (15.72) | 44.88 (20.46) | <0.001 |
|  | 18~102 | 18~88 |  |
| Gender | 1085 (40.8%) | 1112 (68.2%) | <0.001 |

**Appendix E4**
Visualization of all 101 GIS hip ROI detection results – a website link will be available once the manuscript has been accepted for publication.